\documentclass[11pt,preprint]{aastex}
 
\shorttitle{Proper Motions in Lanning UV Sources}
\shortauthors{Lanning \& L\'epine}

\begin{document}

\title{Proper Motions of Faint UV-Bright Sources in the Sandage
  Two-color Survey of the Galactic Plane}
                                   
\author{Howard H. Lanning\altaffilmark{1} and S\'ebastien L\'epine\altaffilmark{2}}
\altaffiltext{1}{National Optical Astronomy Observatory, 950 N. Cherry Ave.,
       Tucson, AZ 85719, lanning@noao.edu}
\altaffiltext{2}{Department of Astrophysics, Division of Physical
  Sciences, American Museum of Natural History, New York, NY 10024, lepine@amnh.org}

\begin{abstract}

Proper motions with values $\geq10$ mas yr$^{-1}$ or $\leq-10$ mas yr$^{-1}$ have been extracted
from the USNO-B1.0 and Tycho II catalogues for all Lanning UV-bright sources
identified in the Sandage Two-color Survey of the Galactic Plane and presented
in Papers I-VI. Of the 572 sources examined, we find at least 213 which
exhibit a significantly large proper motion. Based on the location of
the sources in a reduced proper motion diagram, we demonstrate that
about two thirds of the high proper motion sources are likely or very
likely to be heretofore unidentified white dwarfs.
\end{abstract}

\keywords{stars: white dwarfs, early-type---astrometry:---surveys:}

\section{Introduction}

To date, seven papers have been published providing listings and
finding charts for faint UV-bright sources identified on plates from
the Sandage two-color survey of the Galactic Plane. The most
recently published paper, Paper VII (\citet{lan04}), included proper
motions for many of the sources found using information extracted from
the USNO-B1.0 catalog \citep{mon03}. The combination of the
magnitude, color and proper motion values available for these sources
provides important additional data that may lead to the detection of
previously unidentified white dwarfs. The USNO-B1.0 catalogue was not
available prior to publication of Paper VII. Therefore, it was
considered worthwhile to go back and examine all previously identified
Lanning UV-bright sources in hopes of finding additional evidence to
identify the best candidates for new white dwarfs in the published
catalogues.

As noted in the previous papers devoted to this survey project, the
Sandage two-color survey consists of more than 100 plates measuring
6.6 degrees on a side and centered on the Galactic Plane. Plates were
double-exposed with one image taken in the UV and the second image in
the Blue separated by 12 arcseconds. While the original goal of the
Sandage project was to identify the optical counterparts to X-ray
sources found by the {\em Uhuru} X-ray satellite, the UV-bright
sources found during this analysis do not necessarily correspond to
previously discovered X-ray sources. They do represent a
comprehensive survey of the entire 43 square degree field on each
Schmidt plate examined.

\section{Lanning sources with recorded proper motions}

The USNO-B1.0 catalog contains data obtained from scans of 7435 
Schmidt plates taken for various surveys during a 50 year timespan 
\citep{mon03}. Proper motions tabulated in the catalog were determined 
from observations at all available epochs on the scanned plates. The
USNO-B1.0 also provides optical blue magnitudes from the IIIaJ plates
(B$_J$), optical red magnitudes from IIIaF plates (R$_F$), and
infrared magnitudes from the IVN plates (I$_N$).

The USNO-B1.0 catalog was searched for possible high proper motion
counterparts to the Lanning sources. Objects with a recorded USNO-B1.0
proper motion larger than 10 mas yr$^{-1}$ and located within
10$\arcsec$ of the Lanning sources were selected as possible
counterparts. The positions of the proper motion objects were
extrapolated back to the mean epoch of the Sandage two-color survey
plates (circa 1970), prior to the match. A total of 213 possible high
proper motion counterparts were found. Figure 1 shows the difference
in Right Ascension and Declination between these and their
corresponding Lanning sources. All possible high proper motion
USNOB-1.0 counterparts are found to be within 5$\arcsec$ of the quoted
positions of the Lanning sources, with 194 found within only
2$\arcsec$. This excellent agreement is conclusive evidence that the
USNO-B1.0 proper motion stars are true counterparts to the Lanning
sources.

\begin{figure}
\epsscale{1}
\plotone{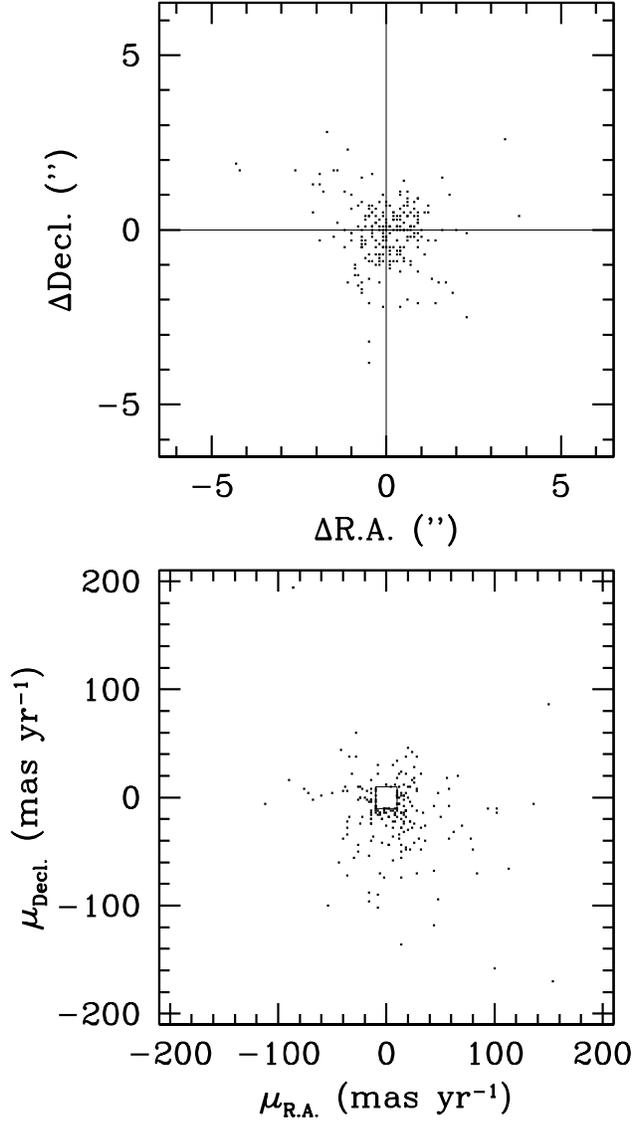}
\caption{Top: difference in the recorded angular positions of Lanning
 UV-bright sources and the recorded positions of their high
 proper motion USNO-B1.0 counterparts. The differences are within the
 combined measurement errors of the Lanning and USNO-B1.0
 catalogs. Bottom: distribution in the recorded proper motions of the
 USNO-B-1.0 sources. Only sources with proper motions
 $\geq10$ mas yr$^{-1}$ or $\leq-10$ mas yr$^{-1}$ in
 either R.A. or Decl. were considered in the analysis.}
\end{figure}

The UV sources thus found to exhibit a large proper motion are listed
in Table I. Source numbers in the table follow the convention
established by the SIMBAD
database\footnote{http://simbad.u-strasbg.fr/Simbad} for Paper I and
have been sorted in numerical order. The format of Table I consists of
1) the Lanning source number, 2) the Sandage Plate Id (plate center in
galactic coords.), 3) Right Ascension (equinox J2000), 4) Declination
(equinox J2000), 5) estimated photographic blue magnitude, 6)
estimated U-B color difference, 7) proper motion in right ascension
$\mu$RA, 8) proper motion in declination $\mu$Decl, 9) photographic
blue (B$_{J}$) magnitude from USNO-B1.0, 10) photographic red
(R$_{F}$) magnitude from USNO-B1.0, 11) photographic infrared (I$_{N}$)
magnitude from USNO-B1.0, 12) the spectral type or estimated spectral
class \--- see \S3 below, and 13) general notes, including
identifications with Tycho-2 stars or previously known objects.

Note that for 15 of the sources, the USNO-B1.0 does not report a value
for the I$_{N}$ magnitude. For those stars, we estimated a rough
magnitude from a direct visual inspection of Digitized Sky Survey
scans.

Entries for Right Ascension, Declination, $m_B$ magnitude and U$-$B
color are taken from the original papers in the series. Corrections to
coordinate errors found following publication have been incorporated here 
and on the web site devoted to the survey (see below). All positions were
measured directly from images extracted from the GSS plates at STScI..
Estimates for the U$-$B color for Paper I \citep{lan73} sources
(Nos. 1-82) were only given in three ranges, strong sources were bluer
than $-$0.8, moderate sources were between $-$0.5 and $-$0.8 and
marginal sources were between 0.0 and $-$0.5 and are represented as
such in the table.

Many of the sources listed in the table were referenced as Tycho-2
entries in the USNO-B1.0 catalog; their Tycho-2 catalog numbers are
noted in the comments. In each case, the proper motion values quoted
in columns 7-8 are from the Tycho-2 catalog \citep{tyc00} instead of
from USNO-B1.0. Proper motions in the Tycho-2 catalog were determined
based upon Tycho-2 observations at a mean epoch of 1991.25 and
observations from 143 transit circle and photographic catalogues
between 1905 and 1991.25, and are more accurate than the USNO-B1.0
values.


\subsection{Notes on individual sources}

\subsubsection{Lanning 16}

This UV source is the star SAO 161359, an O8-9V star which is a
probable member of the Galactic star forming region M17
\citep{HC95}. We found its USNO-B1.0 counterpart to be listed with a
proper motion vector $\mu RA,\mu Decl=-8,+30$ mas yr$^{-1}$. The star is
also listed in the TYCHO-2 catalog (catalog number TYC 6265-1440-1) as
having a proper motion $\mu RA,\mu Decl=-6.1,+30.9$ mas yr$^{-1}$. These
values are, however, inconsistent with an M17 membership. The star
forming region is at an estimated distance of D=$2.5\pm0.3$ Kpc
\citep{C03}; at such a distance, a star with a proper motion of 31 mas
yr$^{-1}$ would have a transverse velocity $\approx360$ km s$^{-1}$
which is inconsistent with a young cluster membership. The large,
quoted proper motion is most likely in error. Indeed, we find that the
star is listed in the Tycho Reference Catalog \citep{H98} with a much
smaller proper motion $\mu RA,\mu Decl=+0.2,-0.4$ mas yr$^{-1}$, a value
which is much more reasonable. We have thus not included this
star in our final list of high proper motion UV sources.

\subsubsection{Lanning 140}

This UV-bright source is one component in a common proper motion pair
with a 7$\arcsec$ angular separation. The Lanning source is the west
component of the pair. The single object listed in the USNO-B1.0
catalog appears to be a compound of the two stars. While the two
components have about the same brightness on the B$_{J}$ plate, the
Lanning source is significantly fainter on the I$_{N}$ plate. The
magnitudes quoted in Table 1 were estimated visually from Digitized
Sky Survey scans. The pair is most likely a main sequence + white
dwarf binary, with Lanning 140 being the white dwarf member. 

\subsubsection{Lanning 526}

The quoted USNOB-1.0 proper motion for this source is $\mu RA$=$-$30 mas yr$^{-1}$, $\mu Decl$=+20 mas yr$^{-1}$.
However, an examination of
Digitized Sky Survey scans shows that this proper motion is most
certainly wrong, as the star is clearly seen moving in the southeast
direction, and with a much larger proper motion than quoted in
USNO-B1.0. A more accurate proper motion measurement performed with
the SUPERBLINK software \citep{LS05} indicates a proper motion 
$\mu RA$=+113 mas yr$^{-1}$, $\mu Decl$=$-$66 mas yr$^{-1}$, which we adopt
in place of the quoted USNOB-1.0 value.

\section{New white dwarfs}

Many of the the Lanning sources, with their large UV excess, are blue
stars with large effective temperatures. Stars with large proper
motions tend to be relatively nearby, which means that any relatively
faint Lanning source with a significant proper motion is most likely
to be a white dwarf. Brighter Lanning sources with moderate proper
motions, on the other hand, are most likely to be more distant main
sequence stars. 

\begin{figure}
\epsscale{1}
\plotone{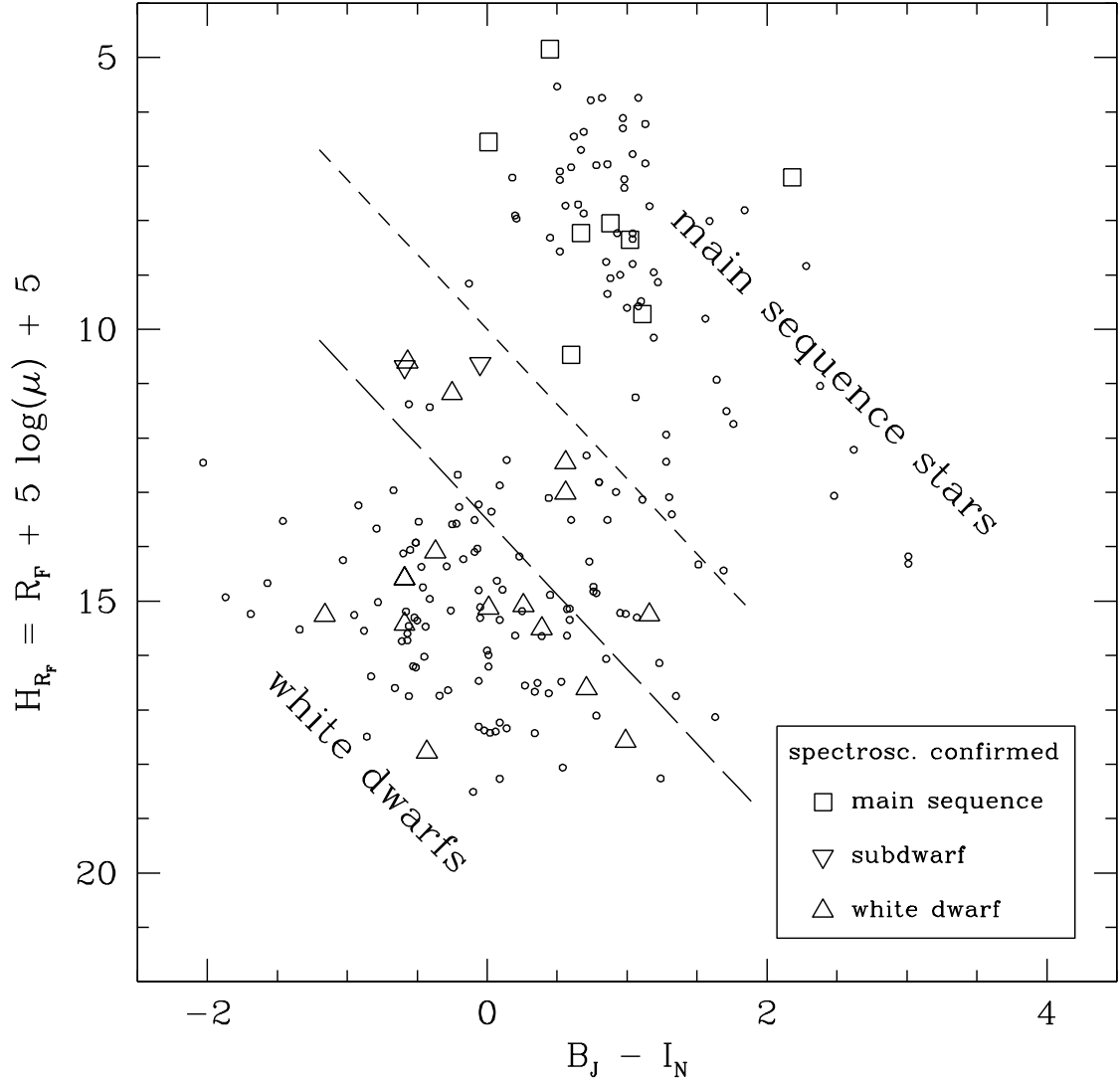}
\caption{Reduced proper motion diagram of the UV-bright Lanning
  sources with large proper motions. The sources are distributed
  around two separate loci, depending on their spectral class: white
  dwarfs populate the lower left of the diagram, main sequence stars
  are found above and to the right of the white dwarf locus. Stars
  located in between the two main groups are possibly subdwarfs. This
  segregation is confirmed by the location of stars which have formal
  spectroscopic classifications (large symbols, see legend). The two
  dashed lines attempt to broadly deliniate regions where Lanning
  sources have a high (bottom), moderate (center), and low probability
  (top) of being white dwarfs.
}
\end{figure}

It is possible to identify the sources which are most likely to be
white dwarfs with the use of a reduced proper motion diagram. We use
the photographic magnitudes of the USNO-B1.0 counterparts (B$_J$, R$_F$,
I$_{N}$) and their recorded proper motion ($\mu$), to build a diagram
of the reduced proper motion H$_{R_F}$=R$_F$+5$\log{\mu}$+5 versus the
B$_J-$I$_{N}$ color. The resulting diagram is shown in Figure 2. The
reduced proper motion diagram is analogous to the standard color
magnitude diagram in that white dwarfs tend to populate the lower left
side of the plot, corresponding to a large H$_{R_F}$ and a blue
B$_J-$I$_{N}$ color; the main sequence stars generally populate a
region above and to the right of the white dwarfs, and subdwarf stars
are generally found in between \citep{SG02,LRS03}. This pattern
occurs because H$_{R_F}$ is correlated with the absolute magnitude of
the source.

The location in the reduced proper motion diagram of sources with
known spectral types clearly demonstrates the separation between main
sequence stars and white dwarfs. We have searched for counterparts of
the Lanning sources in the Catalog of Spectroscopically Identified
White Dwarfs \citep{MS03} and found 16 matches. These are shown as
open triangles in Figure 2. We also searched the Simbad database for
sources with published spectral types; we found 8 stars with main
sequence spectral types from B to G5; these are denoted with open
square symbols in Fig.2. We also found two stars formally classified
as hot subdwarfs (sdO); these are plotted as inverted open triangles. The
spectral types of these objects are also noted in Table 1.

We separate the reduced proper motion diagram into 3 regions which
correpond to an increasing probability for a source to be a white
dwarf. Stars falling in each group are assigned a spectral class
corresponding to the region where they fall. The first region contains
all stars for which:
\begin{equation}
H_{R_F} < 10.0 + 2.75 ( B_J-I_{N} ) ,
\end{equation}
and which are most probably main sequence stars. To these objects we
assign a spectral class ``ms''. The second region includes all
objects for which:
\begin{equation}
10.0 + 2.75 ( B_J-I_{N} ) < H_{R_F} < 13.5 + 2.75 ( B_J-I_{N} ),
\end{equation}
and which probably contain a mix of white dwarfs and subdwarfs, with
perhaps some main sequence stars as well; these stars are assigned a
spectral class ``wd/sd''. Finally, the third region includes all stars
with:
\begin{equation}
H_{R_F} > 13.5 + 2.75 (B_J-I_{N}) ,
\end{equation}
which are the objects most likely to be white dwarfs, and to which we
assign a spectral class ``wd''. The tentative spectral classes are
given in Table 1 for all stars which do not have formal spectral
classification at this time. These are meant to be used as guides to
future spectroscopic follow-up surveys.


\section{Conclusions}

We have determined that 213 UV-bright, Lanning sources have
counterparts with significant proper motions, as recorded in the
USNO-B1.0 catalog. For a majority of the sources, in particular the
faint blue objects, the large proper motion strongly suggests that they
are white dwarfs. The location of the Lanning sources in the
reduced proper motion diagram confirms this impression, and indicates
which objects are most likely to be white dwarfs. Spectroscopic
follow-up of these sources should uncover a significant number of
formerly uncatalogued white dwarfs in the Solar vicinity.

Information related to this survey including published papers, finding
charts, updated tables, etc. may be viewed on the survey web
site\footnote{See http://www.noao.edu/noao/staff/lanning/uvsurvey/index.html}.

\acknowledgments

The assistance of Dr. Stephen Levine, United States Naval Observatory,
Flagstaff, in extracting the proper motion information via batch
processing was most appreciated. The original photographic survey was
supported in part by the National Aeronautics and Space Administration
under grant NGR 09-140-009. This continuing research project has been
supported by funding from the STScI, which is operated by the
Association of Universities for Research in Astronomy, Inc. under NASA
contract NAS5-26555 and by the NASA ADP contract PO\# S-92513-Z. SL is
supported by the National Science Fundation under NSF grant \#
AST 0607757. SL also acknowledges support from Mr. Hilary Lipsitz, and from
the American Museum of Natural History. This research has made use of
the USNOFS Image and Catalogue Archive operated by the United States
Naval Observatory, Flagstaff
Station\footnote{http://www.nofs.navy.mil/data/fchpix/}. This research
has also made use of the Simbad reference database and of the VizieR
catalogue service, both maintained by the Centre de Donn\'ees
astronomiques de Strasbourg\footnote{http://cdsweb.u-strasbg.fr/}.

\clearpage

\begin{deluxetable}{llrrrrrrrrrrl}
\tabletypesize{\scriptsize}
\rotate
\tablecaption{Proper Motion in Lanning UV-Bright Sources. \label{tbl-1}}
\tablewidth{0pt}
\tablehead{
\colhead{Lanning} &
\colhead{Sandage} &
\colhead{R.A.} &
\colhead{Decl.} &
\colhead{$m_{B}$} &
\colhead{$U-B$} &
\colhead{$\mu$RA } &
\colhead{$\mu$DEC} &
\colhead{B$_J$} &
\colhead{R$_F$} &
\colhead{I$_N$} &
\colhead{Spectral} &
\colhead{Notes} 
\\
\colhead{number} &
\colhead{Plate Id} &
\colhead{(J2000)} &
\colhead{(J2000)} &
\colhead{mag} &
\colhead{mag} &
\colhead{mas/yr} &
\colhead{mas/yr} &
\colhead{mag} &
\colhead{mag} &
\colhead{mag} &
\colhead{type/class} &
\colhead{}
}
\startdata
 004 & 123+6 & 01 26 08.8 & 69 01 55.3 & 16 & $<$-0.8 & -22 & 8 & 15.9 & 16.3 & 16.5 & wd &\\
 005 & 134+0 & 02 13 23.0 & 62 04 43.6 & 20: & $<$-0.8 & 22 & -48 & 16.7 & 16.6 & 17.0 & wd &\\
 012 & 193+0 & 06 09 08.6 & 16 42 16.6 & 19: & $<$-0.8 & 2 & -10 & 17.4 & 17.8 & 17.4 & wd/sd &\\
 013 & 193+0 & 06 19 28.9 & 18 38 01.2 & 19: & $<$-0.8 & -4 & -42 & 17.2 & 16.9 & 17: & wd &\\
 018 & 33.2+0 & 18 47 39.1 & 01 57 39.1 & 12 & $<$-0.8 & -2 & -74 & 12.9 & 13.1 & 12.3 & DA2 &\\
 019 & 53+0 & 19 33 49.9 & 18 52 03.1 & 15.5 & $<$-0.8 & 10 & -10 & 15.4 & 15.7 & 15.8 & wd/sd &\\
 020 & 53+0 & 19 43 31.2 & 18 24 35.2 & 12 & $<$-0.8 & 3.3 & -43.8& 12.1 & 12.5 & 12.7 & sdO &TYC 1606-367-1,ALS 10448\\
 021 & 60.1+0 & 19 44 59.1 & 22 45 49.2 & 16 & $<$-0.8 & -8 & -16 & 15.4 & 16.2 & 17.4 & wd &\\
 023 & 111.1+0 & 22 49 52.4 & 58 34 31.9 & 14 & $<$-0.8 & 8 & -14 & 14.3 & 14.6 & 14.8 & DA1 &\\
 024 & 123+6 & 01 24 05.1 & 69 13 20.3 & 19: &-0.5 to -0.8 & 94 & -10 & 17.0 & 17.2 & 16.3 & wd &\\
 029 & 148-6 & 03 17 51.6 & 53 04 20.0 & 17 &-0.5 to -0.8 & -14 & -16 & 17.4 & 17.6 & 17.5 & wd &\\
 032 & 172+0 & 05 20 00.1 & 36 34 06.7 & 15.7 &-0.5 to -0.8 & -10 & -8 & 15.4 & 15.6 & 15.6 & DA &WD 0516+365\\
 034 & 193+0 & 06 08 23.9 & 20 30 11.1 & 18: &-0.5 to -0.8 & 10 & -6 & 17.6 & 17.1 & 17.5 & wd/sd &\\
 036 & 193+0 & 06 18 45.6 & 15 16 52.2 & 13 &-0.5 to -0.8 & -32 & 22 & 12.7 & 13.1 & 13.3 & wd/sd &\\
 048 & 53.4-6 & 20 03 17.4 & 16 18 44.4 & 12 &-0.5 to -0.8 & -2.2 &-18.3 & 12.5 & 12.6 & 12.7 & ms &TYC 1617-1006-1\\
 050 & 53.4-6 & 20 07 36.5 & 17 42 15.4 & 15.3 &-0.5 to -0.8 & 102 & -10 & 15.2 & 15.0 & 14.9 & DApe & WZ Sge \\
 051 & 96.1+0 & 21 18 56.7 & 54 12 34.3 & 12.5 &-0.5 to -0.8 & -85.4& 195.2& 12.6 & 12.9 & 13.2 & DA3.5 &TYC 3953-480-1\\
 057 & 53.4-6 & 03 10 03.2 & 53 09 24.5 & 17 & 0.0 to -0.5 & 28 & -70 & 16.6 & 16.2 & 17.2 & wd &\\
 062 & 53.4-6 & 03 31 18.1 & 53 03 52.0 & 17 & 0.0 to -0.5 & -18 & -2 & 16.8 & 17.0 & 17.0 & wd/sd &\\
 079 & 18-3 & 18 28 27.7 & -17 33 01.3 & 11 & 0.0 to -0.5 & 12.7& -4.5& 11.7 & 11.6 & 11.5 & ms &TYC 6270-1710-1\\
 082 & 96.1+0 & 21 26 24.9 & 55 13 29.3 & 14 & 0.0 to -0.5 & 268 & 192 & 15.7 & 15.0 & 14.8 & DA4 &\\
 083 & 88.0+6 & 20 11 15.0 & 49 10 37.0 & 20.5 & $-$1.3 & 150 & 86 & 17.6 & 17.3 & 17.8 & wd &\\
 088 & 88.0+6 & 20 17 21.9 & 45 52 50.5 & 20.5 & $-$1.2 & 18 & 22 & 16.6 & 17.4 & 18.2 & wd &\\
 089 & 88.0+6 & 20 17 55.0 & 47 51 43.5 & 20.5 & $-$1.4 & 10 & 34 & 16.2 & 17.5 & 17.9 & wd &\\
 091 & 88.0+6 & 20 24 35.2 & 45 05 28.7 & 18.5 & $-$1.0 & 16 & 2 & 16.7 & 17.5 & 18.1 & wd &\\
 094 & 88.0+6 & 20 32 28.9 & 48 01 48.1 & 19 & $-$0.6 & -40 & -38 & 17.2 & 16.5 & 16: & wd/sd &\\
 097 & 96+6 & 20 40 45.1 & 53 03 46.1 & 20.5 & $-$0.7 & 10 & 18 & 19.0 & 18.6 & 18.0 & wd/sd &\\
 098 & 96+6 & 20 42 20.0 & 51 42 14.4 & 20.5 & -0.5 & 10 & 4 & 17.7 & 17.2 & 17.0 & ms &\\
 100 & 96+6 & 20 43 51.7 & 53 13 25.8 & 21.5 & $-$0.9 & -14 & -24 & 19.4 & 18.6 & 18: & wd &\\
 101 & 88.0+6 & 20 42 51.2 & 47 09 06.8 & 20.5 & $-$1.0 & 6 & -36 & 17.5 & 17.0 & 17.4 & wd &\\
 108 & 96+6 & 20 54 12.3 & 55 04 09.1 & 20 & $-$0.5 & 12 & -6 & 18.7 & 18.5 & 18.5 & wd/sd &\\
 109 & 96+6 & 20 55 21.0 & 50 14 07.4 & 21 & $-$1.0 & -8 & 22 & 18.2 & 18.0 & 17.8 & wd/sd &\\
 112 & 96+6 & 21 00 31.4 & 50 51 19.2 & 15.5 & $-$0.4 & -54 & -100 & 16.0 & 15.0 & 14.9 & DA5 &\\
 113 & 96+6 & 21 02 46.4 & 55 30 18.5 & 20.5 & $-$1.4 & 20 & 10 & 18.8 & 18.3 & 18: & wd &\\
 114 & 96+6 & 21 01 54.5 & 51 38 56.6 & 21 & $-$0.6 & -36 & 6 & 17.9 & 18.2 & 17.9 & wd &\\
 115 & 96+6 & 21 04 35.3 & 55 57 39.4 & 21 & $-$1.2 & -14 & -40 & 18.4 & 17.1 & 17.4 & wd/sd &\\
 116 & 96+6 & 21 04 48.8 & 55 35 15.9 & 8 & $-$0.2 & 12.6& 22.6& 8.2 & 7.8 & 7.7 & Am &TYC 3956-803-1, HD 201033\\
 119 & 96+6 & 21 12 23.0 & 55 55 24.6 & 20.5 & $-$0.5 & -24 & 10 & 16.4 & 16.6 & 17.2 & wd &\\
 120 & 96+6 & 21 14 05.7 & 49 51 17.0 & 19 & moderate & 16 & -8 & 17.0 & 17.9 & 17.6 & wd &\\
 122 & 96+6 & 21 18 56.3 & 54 12 45.4 & 13 & $-$0.6 & -85.4& 195.2& 12.6 & 12.9 & 13.2 & DA3.5 &TYC 3953-480-1\\
 123 & 117+6 & 23 26 58.9 & 71 00 19.1 & 20.5 & -1.5 & -10 & 8 & 17.0 & 16.4 & 15.7 & ms &\\
 126 & 117+6 & 23 39 10.5 & 66 51 45.6 & 19.5 & $-$0.8 & 44 & -8 & 16.8 & 17.3 & 18.2 & wd &\\
 127 & 117+6 & 23 39 45.5 & 67 46 54.9 & 20.5 & $-$1.0 & 16 & -2 & 22.0 & 18.1 & 19.0 & ms &\\
 128 & 117+6 & 23 47 07.2 & 65 34 18.7 & 20 & -0.6 & 10 & 6 & 18.6 & 16.9 & 16.0 & ms &\\
 131 & 117+6 & 00 01 50.6 & 66 07 01.4 & 21.5 & $-$0.3 & -26 & -2 & 18.6 & 18.0 & 18.7 & wd &\\
 133 & 117+6 & 00 12 24.8 & 65 28 53.6 & 21 & $-$0.6 & -36 & -28 & 17.6 & 18.1 & 18.4 & wd &\\
 136 & 134+6 & 02 01 31.7 & 69 42 52.6 & 17.8 & $-$0.8 & -14 & 4 & 18.0 & 18.4 & 19.0 & wd &\\
 138 & 134+6 & 02 31 45.1 & 69 23 32.0 & 11.0 & $-$0.3 & 1.3& -25.9& 11.1 & 10.6 & 10.5 & ms &TYC 4316-2117-1\\
 140 & 134+6 & 02 23 57.1 & 68 49 08.7 & 16.0 & $-$0.3 & -44 & -60 & 15.5 & 16.0 & 16.0 & wd &\\
 142 & 134+6 & 02 28 50.0 & 68 35 37.9 & 17.2 & $-$0.7 & 14 & -50 & 17.6 & 18.0 & 18.3 & wd &\\
 143 & 134+6 & 03 03 27.8 & 68 29 54.1 & 16.5 & $-$0.3 & 50 & -18 & 16.8 & 16.8 & 17.4 & wd &\\
 144 & 134+6 & 02 30 35.7 & 68 29 36.5 & 16.7 & $-$0.5 & 18 & -12 & 18.3 & 18.5 & 17.8 & wd/sd &\\
 146 & 134+6 & 02 38 07.8 & 66 57 36.9 & 17.8 & $-$0.5 & -6 & -16 & 19.4 & 18.7 & 18.7 & wd/sd &\\
 147 & 134+6 & 02 05 37.9 & 65 59 36.0 & 18.0 & $-$0.3 & 14 & 24 & 20.1 & 18.9 & 18.9 & wd/sd &\\
 148 & 134+6 & 02 59 20.1 & 65 05 45.1 & 10.5 & $-$0.2 & 59.2& 9.2& 11.1 & 10.5 & 10.3 & ms &TYC 4056-913-1\\
 150 & 134+6 & 02 15 34.6 & 64 53 22.2 & 16.0 & $-$0.5 & 44 & -68 & 17.8 & 17.0 & 17.5 & wd &\\
 151 & 134+6 & 03 00 56.7 & 64 10 54.3 & 17.5 & $-$0.6 & -14 & -14 & 18.0 & 18.9 & 18: & wd &\\
 153 & 134+6 & 02 29 19.2 & 64 03 30.3 & 16.5 & $-$0.4 & -10 & -14 & 17.7 & 16.9 & 16.4 & ms &\\
 155 & 128+6 & 01 18 55.7 & 71 20 11.4 & 18: & -0.3 & 10 & 2 & 17.1 & 16.2 & 16.0 & ms &\\
 157 & 128+6 & 02 18 21.4 & 70 47 48.1 & 18.0 & $-$0.8 & 6 & -12 & 17.8 & 17.6 & 18.8 & wd &\\
 158 & 128+6 & 02 19 06.3 & 70 08 39.8 & 16.5 & $-$1.0 & 36 & -40 & 16.6 & 16.6 & 17.1 & wd &\\
 159 & 128+6 & 01 55 10.9 & 69 42 41.3 & 17.0 & $-$0.3 & -112 & -6 & 17.4 & 17.1 & 17.3 & wd &\\
 160 & 128+6 & 01 21 57.1 & 69 24 50.6 & 17.6 & $-$0.4 & 18 & -12 & 18.1 & 18.6 & 19.0 & wd &\\
 161 & 128+6 & 01 19 41.4 & 68 51 10.8 & 18.0 & $-$0.5 & -26 & 10 & 17.8 & 17.2 & 16.1 & ms &\\
 163 & 128+6 & 01 33 38.3 & 68 03 32.7 & 16.5 & $-$0.5 & -34 & -18 & 17.6 & 16.8 & 16.9 & wd/sd &\\
 165 & 128+6 & 01 51 09.9 & 67 39 31.3 & 13.5 & $-$0.6 & 136 & -6 & 14.1 & 14.6 & 15.2 & DA2 &\\
 166 & 128+6 & 01 56 06.2 & 67 14 45.5 & 18: & $-$0.6 & -22 & -18 & 16.6 & 17.7 & 18.5 & wd &\\
 167 & 128+6 & 02 18 18.9 & 67 02 25.4 & 17.6 & $-$0.8 & -14 & -4 & 18.0 & 18.1 & 18.5 & wd &\\
 169 & 128+6 & 01 32 42.9 & 66 35 47.3 & 16.5 & $-$0.2 & 32 & -28 & 17.1 & 16.1 & 16.4 & wd/sd &\\
 171 & 128+6 & 02 14 52.9 & 65 57 59.0 & 17.0 & $-$0.5 & 32 & -48 & 18.4 & 17.9 & 18.1 & wd &\\
 174 & 148+6 & 03 41 17.1 & 62 42 03.1 & 16.8 & $-$0.7 & 2 & -18 & 17.1 & 17.2 & 17.6 & wd &\\
 175 & 148+6 & 03 59 51.1 & 62 29 01.3 & 19.0 & $-$0.6 & 30 & -38 & 18.6 & 18.6 & 18: & wd &\\
 176 & 148+6 & 03 56 51.0 & 62 26 20.0 & 12.5 & $-$0.2 & 10 & -18 & 12.6 & 11.2 & 10.8 & ms &\\
 177 & 148+6 & 03 51 15.7 & 61 52 46.4 & 17.0 & $-$0.5 & 84 & -70 & 16.8 & 17.3 & 17.6 & wd &\\
 183 & 148+6 & 04 23 08.3 & 60 02 39.8 & 11.4 & $-$0.4 & -6.7& -11.8& 11.7 & 11.3 & 11.1 & ms &TYC 4065-34-1\\
 184 & 148+6 & 04 20 00.2 & 60 05 21.8 & 11.0 & $-$0.4 & -25.6& -49.5& 11.2 & 10.4 & 10.0 & ms &TYC 4065-2-1\\
 188 & 148+6 & 04 16 02.5 & 59 44 02.4 & 17.5 & $-$0.6 & 6 & -22 & 17.8 & 18.2 & 18.5 & wd &\\
 191 & 148+6 & 04 04 23.8 & 58 58 36.2 & 16.8 & $-$0.3 & -14 & 8 & 17.6 & 17.5 & 17.8 & wd &\\
 193 & 148+6 & 04 22 46.1 & 58 22 11.4 & 16.5 & $-$0.4 & 6 & -28 & 16.8 & 16.1 & 15.5 & ms &\\
 197 & 123-6 & 00 53 01.7 & 59 59 42.6 & 15.0 & $-$0.6 & -76 & 8 & 16.4 & 16.2 & 15.8 & wd &\\
 209 & 123-6 & 00 54 23.3 & 59 27 27.8 & 12.0 & $-$0.2 & 25.6& -7.6& 11.6 & 11.3 & 11.2 & ms &TYC 3680-887-1\\
 212 & 123-6 & 00 54 33.9 & 59 16 49.3 & 12.3 & $-$0.2 & -14.9& -3.7& 11.4 & 10.7 & 10.4 & ms &TYC 3680-539-1\\
 225 & 123-6 & 00 43 10.5 & 58 14 20.8 & 18.5 & $-$0.6 & 16 & -16 & 18.5 & 18.0 & 17.8 & wd/sd &\\
 226 & 123-6 & 01 12 23.8 & 58 07 50.5 & 10.5 & $-$0.5 & -67.6& -1.4& 9.9 & 9.2 & 8.9 & G0 &TYC 3681-121-1,HD 236653\\
 234 & 123-6 & 01 12 50.9 & 57 16 43.4 & 11.5 & $-$0.3 & -19.7& 0.3& 10.3 & 9.6 & 9.3 & ms &TYC 3677-1423-1\\
 239 & 123-6 & 01 09 17.2 & 57 07 27.6 & 12.5 & $-$0.3 & 4.2& -11.7& 11.7 & 11.2 & 11.0 & ms &TYC 3677-442-1\\
 241 & 123-6 & 01 13 07.1 & 57 05 58.3 & 11.8 & $-$0.4 & -12.3& -18.2& 10.8 & 10.0 & 9.7 & ms &TYC 3677-1939-1\\
 245 & 123-6 & 01 12 26.8 & 56 39 59.7 & 12.2 & $-$0.3 & 49.8& 5.1& 11.9 & 11.2 & 10.9 & ms &TYC 3677-1474-1\\
 247 & 123-6 & 01 10 29.1 & 56 37 40.2 & 10.8 & $-$0.5 & 11.8& -4.7& 10.6 & 10.2 & 10.1 & ms &TYC 3677-2235-1\\
 248 & 123-6 & 00 53 28.9 & 56 38 36.7 & 11.2 & $-$0.2 & -4.3& -12.5& 10.7 & 10.6 & 10.6 & B &TYC 3676-2052-1\\
 253 & 123-6 & 00 33 40.7 & 55 51 47.0 & 14.8 & $-$0.6 & 14 & -14 & 16.6 & 16.5 & 15.6 & wd/sd &\\
 260 & 123-6 & 01 09 20.6 & 55 10 43.3 & 12.5 & $-$0.3 & 9.4& -15.6& 11.3 & 10.7 & 10.4 & ms &TYC 3673-1035-1\\
 272 & 141+0 & 02 57 45.1 & 60 34 27.4 & 14.0 & $-$0.5 & 10 & -12 & 15.4 & 15.4 & 15.9 & wd/sd &\\
 273 & 141+0 & 03 08 18.4 & 60 35 30.1 & 17.0 & $-$0.6 & 14 & -74 & 17.6 & 17.3 & 17.2 & wd &\\
 280 & 141+0 & 02 48 41.3 & 59 16 12.4 & 16.5 & $-$0.7 & 44 & -118 & 16.3 & 16.1 & 16.6 & wd &\\
 284 & 141+0 & 03 04 17.9 & 58 44 05.2 & 17.0 & $-$1.2 & -36 & -22 & 16.5 & 16.6 & 17.0 & wd &\\
 288 & 141+0 & 03 07 17.2 & 57 50 29.5 & 18.5 & $-$0.6 & 26 & -24 & 17.4 & 17.8 & 18.3 & wd &\\
 290 & 141+0 & 03 01 09.0 & 57 39 56.4 & 19.5 & $-$0.5 & 78 & -38 & 18.5 & 18.4 & 18.0 & wd &\\
 291 & 141+0 & 03 03 31.9 & 57 19 14.8 & 19.5 & $-$0.8 & 12 & 0 & 17.9 & 17.0 & 16.6 & ms &\\
 293 & 141+0 & 03 09 21.8 & 57 05 21.5 & 19.5 & $-$0.5 & 18 & -12 & 18.4 & 18.7 & 17.8 & wd/sd &\\
 294 & 141+0 & 03 21 05.6 & 56 54 43.5 & 19.0 & -0.6 & 10 & -10 & 18.6 & 18.2 & 18: & wd &\\
 295 & 141+0 & 03 06 16.1 & 56 51 43.3 & 18.5 & $-$0.6 & 2 & -18 & 17.7 & 18.1 & 18.0 & wd &\\
 296 & 141+0 & 03 04 13.4 & 56 15 55.9 & 19.5 & $-$0.5 & 0 & 0 & 18.4 & 18.3 & 18: & wd &\\
 297 & 141+0 & 03 10 05.5 & 56 03 18.8 & 19.0 & $-$0.6 & 22 & -22 & 18.4 & 18.2 & 18.0 & wd &\\
 298 & 141+0 & 02 55 16.3 & 55 53 25.2 & 17.5 & $-$0.4 & 20 & 46 & 17.5 & 16.8 & 16.4 & wd/sd &\\
 300 & 141+0 & 03 25 21.9 & 55 00 08.9 & 16.5 & $-$0.5 & -8 & -90 & 16.9 & 16.7 & 17.0 & wd &\\
 303 & 141+0 & 03 00 59.1 & 54 32 42.3 & 17.0 & $-$0.5 & 154 & -170 & 17.2 & 16.5 & 16.0 & wd &\\
 311 & 148+0 & 03 47 21.0 & 55 51 52.1 & 17.0 & $-$0.4 & -16 & -88 & 16.6 & 17.0 & 17.2 & wd &\\
 313 & 148+0 & 04 08 11.1 & 55 50 12.8 & 12.0 & $-$0.3 & 0.1& 12.0& 12.7 & 12.5 & 12.5 & ms &TYC 3722-650-1\\
 316 & 148+0 & 03 39 21.7 & 55 25 29.6 & 18.5 & -0.6 & 10 & 0 & 18.1 & 18.2 & 18.1 & wd/sd &\\
 318 & 148+0 & 04 11 26.4 & 55 11 39.8 & 11.0 & $-$0.2 & 16.4& -30.1& 11.3 & 10.6 & 10.3 & ms &TYC 3722-789-1\\
 321 & 148+0 & 03 57 42.3 & 54 04 22.6 & 18.0 & $-$0.5 & 18 & -48 & 16.9 & 16.9 & 17.3 & wd &\\
 328 & 148+0 & 03 36 41.2 & 53 02 30.3 & 12.0 & $-$0.3 & 14.2& -26.7& 11.5 & 11.1 & 10.9 & ms &TYC 3716-334-1\\
 337 & 148+0 & 03 39 58.3 & 52 09 53.9 & 17.5 & $-$0.8 & 24 & 42 & 16.8 & 17.3 & 17.4 & wd &\\
 339 & 148+0 & 03 52 06.9 & 51 40 03.5 & 12.0 & $-$0.1 & -33.4& 39.4& 11.1 & 10.5 & 10.2 & ms &TYC 3338-1104-1\\
 351 & 148+0 & 04 06 07.1 & 54 31 33.9 & 14.5 & $-$0.6 & -90 & 16 & 15.6 & 15.5 & 15.5 & wd &\\
 353 & 80.4+0 & 20 54 27.3 & 42 42 44.2 & 18.0 & -0.3 & -2 & -10 & 18.3 & 17.8 & 17.5 & wd/sd &\\
 356 & 80.4+0 & 20 24 33.0 & 41 23 37.2 & 17.0 & $-$0.4 & 16 & 12 & 17.3 & 17.1 & 17.6 & wd &\\
 358 & 80.4+0 & 20 33 52.7 & 40 56 47.3 & 18.3 & -0.3 & 10 & 2 & 18.8 & 18.3 & 18.7 & wd/sd &\\
 359 & 80.4+0 & 20 50 29.7 & 40 20 33.5 & 12.0 & $-$0.3 & 15.9& -55.1& 10.7 & 10.0 & 9.7 & ms &TYC 3171-452-1\\
 360 & 80.4+0 & 20 38 22.6 & 40 17 34.4 & 18.5 & $-$0.2 & 14 & 4 & 18.6 & 18.3 & 18.7 & wd &\\
 361 & 80.4+0 & 20 48 08.3 & 39 51 38.3 & 14.0 & $-$0.9 & -36 & -34 & 14.8 & 14.5 & 14.2 & DA &\\
 364 & 80.4-6 & 20 56 37.2 & 43 13 26.3 & 17.0 & $-$0.2 & 2 & 10 & 17.8 & 16.5 & 15.9 & wd &\\
 366 & 80.4-6 & 20 54 27.0 & 42 42 47.7 & 18.0 & -0.4 & -2 & -10 & 18.3 & 17.8 & 17.5 & wd/sd &\\
 367 & 80.4-6 & 21 04 02.6 & 42 47 03.5 & 15.5 & $-$0.3 & 32 & 22 & 16.8 & 16.7 & 16.7 & wd &\\
 369 & 80.4-6 & 21 23 12.8 & 42 01 52.4 & 10.6 & $-$0.3 & -11.6& -35.3& 11.1 & 10.3 & 10.0 & ms &TYC 3190-1956-1\\
 370 & 80.4-6 & 21 24 19.1 & 41 58 41.3 & 10.7 & $-$0.4 & 14.3& 3.5& 11.1 & 10.7 & 10.5 & ms &TYC 3190-1682-1\\
 371 & 80.4-6 & 21 23 28.8 & 41 46 33.3 & 10.5 & $-$0.3 & 17.2& 2.0& 10.8 & 10.3 & 10.1 & ms &TYC 3190-2428-1\\
 372 & 80.4-6 & 20 55 47.9 & 41 37 35.4 & 11.8 & $-$0.3 & 16 & -14 & 12.5 & 11.4 & 10.9 & ms &\\
 374 & 80.4-6 & 21 26 24.0 & 41 14 20.0 & 10.8 & $-$0.3 & -12.5& -10.5& 12.0 & 11.5 & 11.3 & ms &TYC 3186-1289 1\\
 375 & 80.4-6 & 21 25 08.0 & 40 53 16.9 & 10.5 & $-$0.2 & -0.8& -10.4& 10.9 & 10.4 & 10.1 & ms &TYC 318-1900-1\\
 377 & 80.4-6 & 20 54 36.3 & 40 33 18.8 & 10.8 & $-$0.3 & -9.0& -10.1& 11.5 & 11.1 & 10.9 & ms &TYC 3171-685-1\\
 378 & 80.4-6 & 20 56 34.2 & 40 19 09.9 & 10.7 & $-$0.3 & 27.7& 29.6& 11.0 & 10.3 & 10.1 & ms &TYC 3171-1120-1\\
 379 & 80.4-6 & 21 23 31.8 & 40 16 05.9 & 10.8 & $-$0.2 & -14.8& -52.9& 12.2 & 11.1 & 10.6 & ms &TYC 3186-2361-1\\
 380 & 80.4-6 & 21 24 46.9 & 39 42 29.9 & 10.6 & $-$0.2 & 5.4& 10.1& 11.3 & 10.6 & 10.3 & ms &TYC 3186-1551-1\\
 390 & 80.4-6 & 21 11 17.7 & 38 43 34.7 & 11.0 & $-$0.3 & 9.1& -15.6& 11.4 & 11.0 & 10.8 & ms &TYC 3169-2148-1\\
 392 & 80.4-6 & 21 13 40.5 & 38 13 02.8 & 17.5 & $-$0.6 & 12 & 10 & 17.4 & 17.5 & 17.5 & wd/sd &\\
 398 & 100+4 & 22 03 47.1 & 60 56 37.6 & 10.8 & $-$0.1 & -0.5& -12.3& 11.3 & 10.5 & 10.1 & ms &TYC 4263-1549-1\\
 401 & 100+4 & 21 16 59.3 & 60 40 30.3 & 11.6 & $-$0.1 & 18 & 38 & 13.7 & 12.8 & 12.0 & ms &\\
 402 & 100+4 & 21 29 15.4 & 60 52 45.9 & 19.0 & $-$0.5 & 28 & 8 & 18.4 & 17.8 & 17.8 & wd/sd &\\
 403 & 100+4 & 21 33 14.4 & 60 48 30.4 & 17.3 & $-$0.6 & 24 & 16 & 17.7 & 17.7 & 17: & wd/sd &\\
 406 & 100+4 & 21 24 44.9 & 60 35 17.5 & 12.4 & $-$0.1 & 8 & 12 & 14.2 & 13.0 & 11.9 & ms &\\
 407 & 100+4 & 21 19 48.2 & 60 21 01.3 & 18.8 & $-$0.8 & 28 & 38 & 16.9 & 16.6 & 17.3 & wd &\\
 413 & 100+4 & 22 04 38.4 & 59 59 43.0 & 17.8 & -0.7 & -10 & -10 & 17.3 & 17.4 & 16.9 & wd/sd &\\
 414 & 100+4 & 21 40 10.9 & 59 55 38.3 & 11.2 & $-$0.1 & -13.6& -23.5& 11.5 & 11.0 & 10.8 & F8 &TYC 3979-143-1, Trumpler 37 1790\\
 416 & 100+4 & 21 28 10.9 & 59 27 13.4 & 19.0 & -1.0 & -10 & -10 & 18.4 & 17.3 & 15.9 & ms &\\
 417 & 100+4 & 21 38 53.1 & 59 29 08.6 & 18.5 & $-$0.6 & 22 & -4 & 18.5 & 18.3 & 18: & wd &\\
 419 & 100+4 & 21 58 39.5 & 59 21 57.8 & 10.5 & $-$0.2 & 29.8& 7.0& 11.3 & 10.4 & 10.1 & ms &TYC 3980-719-1\\
 421 & 100+4 & 22 02 43.0 & 58 44 55.1 & 10.8 & $-$0.2 & 27.2& 18.5& 11.9 & 11.3 & 11.0 & ms &TYC 3981-5-1\\
 422 & 100+4 & 21 43 56.4 & 58 56 35.9 & 18.8 & $-$0.8 & 20 & 0 & 17.7 & 17.5 & 17.8 & wd &\\
 423 & 100+4 & 21 42 14.7 & 58 54 01.7 & 16.5 & $-$0.4 & -14 & -2 & 16.5 & 15.3 & 14.2 & ms &\\
 425 & 100+4 & 21 17 48.3 & 58 43 33.5 & 10.9 & $-$0.2 & -34.8& -71.8& 11.4 & 10.9 & 10.8 & F8 &TYC 3965-1604-1, Trumpler 37 1196\\
 427 & 100+4 & 21 21 55.9 & 58 37 47.3 & 11.1 & $-$0.1 & 56.8& 19.7& 11.7 & 10.9 & 10.6 & G2 &TYC 3978-1304-1, Trumpler 37 1196\\
 430 & 100+4 & 21 20 42.1 & 58 19 25.4 & 17.0 & $-$0.5 & 14 & -136 & 16.5 & 15.9 & 15.8 & DA6 &\\
 432 & 100+4 & 21 58 32.5 & 58 04 33.4 & 18.5 & $-$0.6 & -6 & -70 & 17.4 & 17.2 & 16.9 & wd &\\
 435 & 100+4 & 21 33 29.0 & 58 11 17.2 & 17.5 & $-$0.7 & 0 & 12 & 17.2 & 17.3 & 17.4 & wd/sd &\\
 439 & 100+4 & 21 28 22.9 & 57 38 04.7 & 19.6 & $-$0.6 & 26 & 14 & 18.5 & 17.0 & 15.5 & ms &\\
 444 & 100+4 & 21 35 20.5 & 56 10 49.5 & 17.0 & $-$0.9 & 66 & 20 & 17.6 & 16.9 & 16.7 & wd/sd &\\
 445 & 100+4 & 21 37 40.3 & 56 10 14.0 & 18.8 & $-$0.8 & 70 & -26 & 18.5 & 17.4 & 17.1 & wd/sd &\\
 448 & 100+4 & 21 39 22.4 & 55 59 32.7 & 11.5 & $-$0.2 & -5.0& -12.6& 12.2 & 12.1 & 12.0 & ms &TYC 3971-1001-1\\
 450 & 100+4 & 22 01 22.5 & 55 34 52.4 & 11.0 & $-$0.2 & 3.7& -42.7& 11.5 & 10.8 & 10.6 & ms &TYC 3973-1619-1\\
 454 & 100+4 & 21 41 41.1 & 55 27 50.2 & 11.4 & $-$0.5 & 102.1& -13.9& 10.9 & 10.1 & 9.8 & ms &TYC 3971-891-1\\
 458 & 104.1+6 & 22 45 16.4 & 66 08 13.3 & 18.0 & $-$0.5 & 22 & 10 & 18.8 & 19.0 & 19: & wd &\\
 459 & 104.1+6 & 22 05 38.2 & 62 24 35.7 & 16.0 & $-$0.8 & -38 & 10 & 17.1 & 17.5 & 17.7 & DA &\\
 460 & 141+6 & 03 12 40.1 & 66 44 32.3 & 11.3 & $-$0.1 & -4.0& -19.4& 11.6 & 10.9 & 10.6 & ms &TYC 4061-302-1\\
 461 & 141+6 & 02 54 14.2 & 66 30 35.1 & 17.8 & $-$1.1 & 80 & -48 & 17.5 & 17.6 & 17.2 & wd &\\
 463 & 141+6 & 03 27 08.8 & 66 30 59.3 & 10.4 & $-$0.2 & 58.9& -37.5& 11.1 & 10.3 & 10.0 & ms &TYC 4074-1144-1\\
 465 & 141+6 & 03 11 02.3 & 66 06 04.9 & 10.5 & $-$0.1 & -12.6& -2.0& 10.6 & 10.0 & 9.8 & ms &TYC 4061-89-1\\
 468 & 141+6 & 03 42 22.6 & 65 35 43.1 & 10.2 & $-$0.1 & -59.5& 3.8& 10.9 & 10.1 & 9.7 & ms &TYC 4070-398-1\\
 469 & 141+6 & 03 09 45.0 & 65 39 34.3 & 19.5 & $-$1.0 & 26 & -42 & 18.1 & 18.5 & 19: & wd &\\
 470 & 141+6 & 03 31 53.8 & 65 31 48.1 & 19.0 & $-$0.6 & -28 & 60 & 18.0 & 18.1 & 18.0 & wd &\\
 472 & 141+6 & 03 23 53.9 & 65 08 56.8 & 20.5 & $-$1.0 & 48 & -94 & 18.7 & 18.1 & 18.6 & wd &\\
 476 & 141+6 & 02 53 34.7 & 64 30 37.3 & 10.5 & $-$0.1 & -16.8& 6.8& 10.6 & 9.9 & 9.6 & ms &TYC 4056-704-1\\
 477 & 141+6 & 02 54 56.1 & 64 22 55.3 & 16.8 & $-$0.7 & -14 & -20 & 16.8 & 16.6 & 16.0 & wd/sd &\\
 478 & 141+6 & 03 11 30.4 & 64 25 14.4 & 19.2 & $-$0.4 & 6 & -58 & 18.2 & 18.5 & 18.1 & wd &\\
 480 & 141+6 & 03 29 05.7 & 64 04 42.9 & 11.3 & $-$0.9 & 29.6& -11.2& 13.2 & 13.2 & 13.2 & sdO &TYC 4070-2419-1,ALS 7220\\
 484 & 141+6 & 02 58 15.1 & 63 09 26.3 & 11.5 & $-$0.4 & -29.0& -54.7& 11.3 & 10.6 & 10.2 & ms &TYC 4052-1027-1\\
 486 & 141+6 & 03 24 31.3 & 62 50 53.8 & 15.3 & $-$0.7 & -24 & 0 & 15.8 & 16.1 & 16.5 & wd &\\
 489 & 141+6 & 03 00 37.2 & 62 28 27.9 & 16.8 & $-$0.5 & 6 & 30 & 16.3 & 16.6 & 16.9 & wd &\\
 491 & 141+6 & 03 19 15.4 & 62 05 02.3 & 18.3 & $-$0.7 & 36 & -26 & 17.3 & 17.5 & 17.9 & wd &\\
 494 & 141+6 & 02 57 22.3 & 61 45 09.5 & 19.0 & $-$1.0 & 18 & -18 & 18.0 & 18.3 & 18.1 & wd &\\
 495 & 141+6 & 03 09 31.9 & 61 26 20.9 & 17.8 & $-$0.6 & 58 & -8 & 17.3 & 17.4 & 17.9 & wd &\\
 497 & 141+6 & 03 21 14.3 & 61 05 26.8 & 17.5 & $-$0.9 & -14 & 16 & 17.1 & 16.9 & 16.5 & wd/sd &\\
 499 & 141+6 & 03 18 34.8 & 65 00 03.9 & 16.0 & $-$0.8 & 100 & -158 & 16.4 & 16.4 & 16.8 & DA3 &\\
 500 & 164-6 & 04 23 12.9 & 45 04 51.2 & 17.1 & $-$0.5 & -16 & -96 & 17.6 & 17.5 & 17.6 & wd &\\
 501 & 164-6 & 04 15 52.6 & 45 01 29.8 & 18.5 & $-$0.5 & 24 & -28 & 17.9 & 17.4 & 17.6 & wd &\\
 503 & 164-6 & 04 47 13.4 & 44 24 34.8 & 10.6 & $-$0.2 & 17.4& -21.1& 11.0 & 10.5 & 10.3 & ms &TYC 2905-1267-1\\
 513 & 164-6 & 04 39 22.8 & 42 48 01.8 & 19.0 & $-$0.6 & 18 & -46 & 18.4 & 18.9 & 18.4 & wd &\\
 515 & 164-6 & 04 14 36.9 & 42 22 49.0 & 11.0 & $-$0.4 & 12.6& -37.4& 10.7 & 10.1 & 9.8 & K5 &TYC 2886-1029-1,HD 276118\\
 521 & 164-6 & 04 32 36.4 & 42 06 11.9 & 20.0 & $-$0.5 & 8 & -22 & 18.2 & 17.5 & 16.7 & ms &\\
 525 & 164-6 & 04 45 19.3 & 41 58 40.7 & 10.5 & $-$0.3 & 34 & -14 & 10.2 & 12.3 & 14.8 & ms &\\
 526 & 164-6 & 04 12 03.8 & 41 37 19.3 & 18.5 & $-$0.2 & 113 & -66 & 17.4 & 16.5 & 15.8 & wd/sd &\\
 531 & 164-6 & 04 36 27.6 & 40 47 08.6 & 16.5 & $-$0.5 & -42 & 44 & 16.9 & 16.6 & 16.5 & DA6 &\\
 532 & 164-6 & 04 12 42.9 & 40 41 26.5 & 16.0 & $-$0.7 & 14 & -44 & 16.4 & 16.9 & 17.0 & wd &\\
 533 & 164-6 & 04 31 15.6 & 41 18 11.9 & 13.0 & $-$0.6 & 62 & -32 & 18.0 & 18.1 & 18.1 & wd &\\
 534 & 164-6 & 04 19 51.4 & 40 22 46.8 & 15.5 & $-$0.7 & -24 & -24 & 16.2 & 16.4 & 16.6 & DA &\\
 535 & 164-6 & 04 22 59.8 & 40 16 18.5 & 20.0 & $-$0.7 & 12 & -34 & 19.4 & 19.6 & 19: & wd &\\
 536 & 164-6 & 04 26 28.1 & 40 13 54.2 & 19.0 & $-$0.6 & 22 & -22 & 18.8 & 19.0 & 18.5 & wd &\\
 537 & 164-6 & 04 42 10.0 & 39 59 03.8 & 11.0 & $-$0.3 & -22.5& 0.6& 11.8 & 10.3 & 9.6 & G5 &TYC 2897-262-1,HD 276727\\
 540 & 164-6 & 04 16 39.2 & 40 02 33.9 & 17.0 & $-$1.0 & -72 & 4 & 16.3 & 15.5 & 16.4 & wd &\\
 541 & 164-6 & 04 13 08.9 & 38 56 21.2 & 19.5 & $-$0.6 & 8 & -16 & 19.0 & 19.2 & 19: & wd &\\
 542 & 164-6 & 04 38 39.3 & 41 09 34.4 & 15.0 & $-$0.6 & -8 & -102 & 14.9 & 15.1 & 14.9 & DB3 &\\
 545 & 164+0 & 04 57 38.6 & 45 14 59.1 & 11.0 & $-$0.5 & 14.0& -19.5& 11.0 & 10.3 & 10.0 & ms &TYC 3344-1396-1\\
 548 & 164+0 & 05 07 43.8 & 45 14 05.6 & 10.5 & $-$0.5 & 1.8& -13.5& 11.8 & 11.2 & 11.0 & ms &TYC 3345-2493-1\\
 554 & 164+0 & 04 46 47.3 & 42 15 00.7 & 19.0 & $-$0.5 & 14 & -14 & 18.4 & 19.1 & 18.2 & wd &\\
 556 & 164+0 & 05 13 07.7 & 41 48 50.7 & 19.5 & $-$0.5 & 16 & -24 & 18.6 & 18.6 & 18.6 & wd &\\
 558 & 164+0 & 04 48 43.7 & 41 21 43.9 & 18.5 & $-$0.6 & 14 & -28 & 18.3 & 18.7 & 18.8 & wd &\\
 562 & 164+0 & 05 00 46.1 & 40 34 13.5 & 18.5 & -0.3 & 2 & -10 & 17.8 & 16.7 & 16.1 & ms &\\
 563 & 164+0 & 04 46 12.2 & 40 23 40.2 & 19.0 & -0.3 & 8 & -10 & 17.3 & 16.0 & 15.6 & ms &\\
 567 & 164+0 & 05 15 00.9 & 39 44 11.0 & 19.0 & $-$0.6 & 20 & -30 & 18.7 & 18.8 & 19: & wd &\\
 568 & 164+0 & 05 16 58.5 & 39 34 42.3 & 18.8 & $-$0.6 & -28 & 38 & 18.0 & 18.4 & 18.3 & wd &\\
 570 & 164+0 & 04 55 04.5 & 39 29 29.8 & 18.5 & $-$0.6 & 18 & 4 & 18.0 & 18.0 & 18.4 & wd &\\
 572 & 164+0 & 04 44 50.1 & 39 15 19.7 & 17.0 & $-$0.5 & -26 & -44 & 17.1 & 17.5 & 17.5 & wd &\\
\enddata
\end{deluxetable}

\end{document}